\documentclass[prl,superscriptaddress,showpacs,twocolumn]{revtex4}
\usepackage{graphicx}

\begin{document}

\title{A Magnetic Resonance Realization of Decoherence-Free Quantum Computation}

\author{Jason E. Ollerenshaw}
\author{Daniel A. Lidar}
\affiliation{Department of Chemistry, University of Toronto, Toronto, Ontario,
	Canada M5S 3H6}
\author{Lewis E. Kay}
\email{kay@pound.med.utoronto.ca}
\affiliation{Department of Chemistry, University of Toronto, Toronto, Ontario,
	Canada M5S 3H6}
\affiliation{Department of Biochemistry, University of Toronto, Toronto,
	Ontario, Canada M5S 1A8}
\affiliation{Department of Medical Genetics and Microbiology, University of
	Toronto, Toronto, Ontario, Canada M5S 1A8}

\begin{abstract}
We report the realization, using nuclear magnetic resonance techniques, of the
first quantum computer that reliably executes an algorithm in the presence of
strong decoherence.  The computer is based on a quantum error avoidance code
that protects against a class of multiple-qubit errors.  The code
stores two decoherence-free logical qubits in four noisy physical qubits.
The computer successfully executes Grover's search algorithm in the presence of
arbitrarily strong engineered decoherence.  A control computer with no
decoherence protection consistently fails under the same conditions.
\end{abstract}


\maketitle


A computer that uses the laws of quantum mechanics to store and manipulate
information could in theory perform certain tasks such as
searching~\cite{Gro97} and factoring~\cite{Sho97} with incredible efficiency.
The most critical problem that must be solved to make quantum computing
possible on a useful scale is decoherence, the inevitable process of
entanglement between a quantum computer and its environment.  Decoherence
causes the superposition states that carry information within the computer to
decay rapidly.  Several solutions to the decoherence problem have been
proposed (for a review, see \cite{NC00}).  One technique, quantum error
avoidance, calls for information within the computer to be carried exclusively
by quantum states that are not adversely affected by
decoherence~\cite{ZR98,DG98,LCW98,LBK+01a} (for a review, see~\cite{LW03}).
Here we present the first experimental proof that a nontrivial quantum
computation can be protected against decoherence~\cite{comment1}.  Using
quantum error avoidance, we have constructed a nuclear magnetic resonance
quantum computer~\cite{CFH97,GC97,MBE98} which is unaffected by certain types
of decoherence.  Our computer successfully executes Grover's quantum search
algorithm~\cite{Gro97} in the presence of arbitrarily strong engineered
decoherence.  A control computer with no decoherence protection consistently
fails under the same conditions.

Decoherence is typically characterized by the decay of off-diagonal elements in
a system's density matrix $\rho$.  Formally, decoherence takes the system from
state $\rho_i$ to a state $\rho_f=\sum_dE_d\rho_iE_d^\dagger$ where the Kraus
operators $E_d$ describe transformations that may result from the
system-environment coupling (they satisfy
$\sum_dE_d^\dagger E_d=I=\text{identity}$)~\cite{NC00,K83}.

When the coupling between a quantum system and its environment possesses an
element of symmetry, some of the system's states will be immune to
decoherence~\cite{ZR98,DG98,LCW98,LBK+01a,LW03}.  These states span a
decoherence-free subspace (DFS).  The quantum computer we have constructed
comprises
two decoherence-free logical quantum bits
(qubits~\cite{NC00}), encoded in the DFSs of four noisy physical qubits.  The
code protects against multiple-qubit errors~\cite{LBK+01a}: To satisfy the
symmetry condition for the existence of DFSs, we assume the system-environment
coupling affects certain pairs of qubits rather than affecting each qubit
independently.  Note that this error model is different from the popular
``collective decoherence'' model~\cite{ZR98,DG98,LCW98}.

The multiple qubit errors model is relevant to a number of physical systems
recently used as quantum computers~\cite{LBK+01a}.  For example, the primary
source of decoherence in liquid state NMR is the random modulation of
internuclear dipolar interactions by molecular tumbling.  Under certain
conditions, for example very slow tumbling, the interactions reduce to a
symmetrical multiple qubit error process.  This makes some of a system's
coherences resistant or immune to decoherence (an effect recently
exploited in NMR studies of large proteins~\cite{VHO+03}) and gives rise to
DFSs similar to those used in this work.  However, the object of
this work is to demonstrate DFS protection against multiple qubit errors, not
to demonstrate specific resistance to the natural decoherence processes of
liquid state NMR.  We chose an error model that supports a relatively simple
DFS and that affords us complete control of the decoherence strength, and as such
it is not related to our system's natural decoherence.

Specifically, the code our computer uses resists errors of the
form~\cite{comment2}.
\begin{eqnarray}\label{errors}
E_d & = & a_{d,0}I_1I_2I_3I_4 + a_{d,1}X_1X_2I_3I_4\nonumber\\*
	& & \mbox{} + a_{d,2}I_1I_2X_3X_4 + a_{d,3}X_1X_2X_3X_4,
\end{eqnarray}
where $X_n$ indicates that physical qubit $n\in\{1,2,3,4\}$ is flipped and
$I_n$ indicates that it is unaffected.  There are four DFSs for this set of
errors~\cite{LBK+01a}.  Each DFS is a simultaneous eigenspace of the operators
\{$I_1I_2I_3I_4$, $X_1X_2I_3I_4$, $I_1I_2X_3X_4$, $X_1X_2X_3X_4$\} with
eigenvalues $\pm1$.  The following states are an orthonormal basis for one DFS:
\begin{eqnarray*}
|00\rangle_L^1 & = & \left(|0000\rangle + |1100\rangle + |0011\rangle +
	|1111\rangle\right)/2\\*
|01\rangle_L^1 & = & \left(|1000\rangle + |0100\rangle + |1011\rangle +
	|0111\rangle\right)/2\\*
|10\rangle_L^1 & = & \left(|0001\rangle + |1101\rangle + |0010\rangle +
	|1110\rangle\right)/2\\*
|11\rangle_L^1 & = & \left(|1001\rangle + |0101\rangle + |1010\rangle +
	|0110\rangle\right)/2.
\end{eqnarray*}
These can be used as basis states for two decoherence-free logical qubits and
are labeled as such.  The other three DFSs are related to this DFS by sign
changes.  For example:
\begin{eqnarray*}
|00\rangle_L^2 & = & \left(|0000\rangle + |1100\rangle - |0011\rangle -
	|1111\rangle\right)/2\\*
|00\rangle_L^3 & = & \left(|0000\rangle - |1100\rangle + |0011\rangle -
	|1111\rangle\right)/2\\*
|00\rangle_L^4 & = & \left(|0000\rangle - |1100\rangle - |0011\rangle +
	|1111\rangle\right)/2,
\end{eqnarray*}
where superscript numerals indicate to which DFS a state belongs.  The
remaining basis states of DFSs 2--4 are obtained by applying similar sign
changes to the other states of DFS 1.

The code uses all four of these DFSs in classical parallel.  An arbitrary
state of the two logical qubits, $|\psi\rangle_L = a|00\rangle_L +
b|01\rangle_L + c|10\rangle_L + d|11\rangle_L$ where
$|a|^2 + |b|^2 + |c|^2 + |d|^2 = 1$, is encoded in the density matrix
$\rho_{|\psi\rangle_L}$ describing the four physical qubits according to
\begin{equation}\label{code}
\rho_{|\psi\rangle_L} = |\psi\rangle_L\langle\psi|_L
	= \sum_{i=1}^4c_i|\psi\rangle_L^i\langle\psi|_L^i,
\end{equation}
where $|\psi\rangle_L^i = a|00\rangle_L^i + b|01\rangle_L^i +c|10\rangle_L^i +
d|11\rangle_L^i$ and $c_i = 1/4$.  Note that only the component of the density
matrix that deviates from identity is described above (the identity portion of
$\rho$ is immutable and unobservable during any NMR experiment~\cite{NC00}).

It should be noted that in general a quantum superposition of states from
different DFSs is not decoherence-free.  This is because the DFSs are
eigenspaces of the $E_d$ operators, and for a given $E_d$ each DFS may have a
different eigenvalue.  However, the encoding of Eq.~(\ref{code}) uses the DFSs
in classical superposition only and $\rho_{|\psi\rangle_L}$ is therefore
unaffected by the errors described by Eq.~(\ref{errors}).  It can be shown
that $E_d|\psi\rangle_L^i = A_{d,i}|\psi\rangle_L^i$ where $A_{d,i}$ is a
constant~\cite{LBK+01a}.  It follows that, when the errors of
Eq.~(\ref{errors}) occur:
\begin{eqnarray}
\rho_f & = & \sum_dE_d\rho{}E_d^\dagger
= \sum_d\sum_{i=1}^4c_iE_d|\psi\rangle_L^i\langle\psi|_l^i%
E_d^\dagger\nonumber\\*
& = & \sum_{i=1}^4c_i\left(\sum_dA_{d,i}A_{d,i}^*\right)%
|\psi\rangle_L^i\langle\psi|_L^i
= \sum_{i=1}^4c_i'|\psi\rangle_L^i\langle\psi|_L^i\nonumber
\end{eqnarray}
where $c_i' = c_i\sum_dA_{d,i}A_{d,i}^*$.  Thus the net effect of the errors
is a change in the relative weights $c_i'$ of the different DFSs: The logical
qubit information in each DFS is intact.  For the errors we implement
experimentally, it is always the case that $\sum_dA_{d,i}A_{d,i}^* = 1$ so
that the errors have no effect whatsoever ($c_i' = c_i$ so that
$\rho_f = \rho$).  This simplifies interpretation of the experimental results
but is not essential to the code's performance.

Pulse sequences that perform logic gates on the two encoded qubits were
developed using methods derived in~\cite{LBK+01b} and will be described in
detail in a subsequent publication.
Unfortunately the computer leaves the DFS code during gate sequences.  It has
been shown that this class of DFS codes can function as active
quantum error correction codes against errors occuring during qubit
manipulation~\cite{LBK+01b}, a property that can be used in future
implementations to detect and correct such errors.

Before any computation, the computer's two logical qubits are initialized by
temporal averaging~\cite{KCL98} to the state $|00\rangle_L$.  The code maps
this logical qubit state to the following state of the four physical qubits:
\begin{eqnarray}\label{rhocomp}
\rho_{|00\rangle_L} & = & (|0000\rangle\langle0000| +
	|1100\rangle\langle1100|\nonumber\\*
	& & \mbox{} + |0011\rangle\langle0011| + |1111\rangle\langle1111|) / 4
\end{eqnarray}
This density matrix can be decomposed into a sum of tensor products of the
$2 \times 2$ identity matrix $I$ and the Pauli matrix $Z$.
\begin{eqnarray*}
\rho_{|00\rangle_L} & = & (I_1I_2I_3I_4 + Z_1Z_2I_3I_4\\*
	& & \mbox{} + I_1I_2Z_3Z_4 + Z_1Z_2Z_3Z_4) / 16
\end{eqnarray*}
The $I_1I_2I_3I_4$ term can be neglected and each of the remaining three can
easily be prepared from the system's equilibrium state using standard pulsed
NMR techniques~\cite{EBW87}.  During a quantum computing experiment, the
computation is repeated three times, each time prefaced with a pulse sequence
that prepares a different one of $Z_1Z_2I_3I_4$, $I_1I_2Z_3Z_4$, and
$Z_1Z_2Z_3Z_4$.  The three results are added together and
because computation is a linear quantum operation, the summed result
corresponds to a computation starting from the state $|00\rangle_L$.

Because the code uses states from four DFSs in classical parallel, our ensemble
computer does not use true pseudo-pure states.  We chose this approach to
reduce the number of temporal averaging steps, facilitating a thorough test of
the computer's resistance to decoherence.  In fact, the three temporal
averaging steps we perform are a subset of the fifteen required for a
pseudo-pure state implementation.

\begin{figure*}
\includegraphics{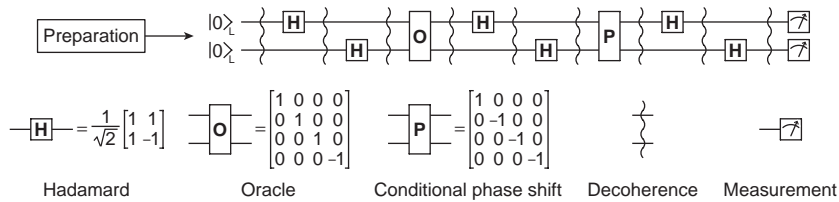}
\caption{\label{grovfig}Running Grover's search algorithm on an error-avoiding
quantum computer.  Time runs from left to right.  Decoherence-free logical
qubits are represented by horizontal lines.  Engineered decoherence is applied
at nine points during the experiment to test the computer's resistance.}
\end{figure*}

We use our computer to perform Grover's quantum search algorithm~\cite{Gro97}
(Fig.~\ref{grovfig}).
The algorithm's purpose is to retrieve, from an unsorted list, the
single item that satisfies a given criterion.
Grover's algorithm is highly efficient, requiring only
$\mathcal{O}(\sqrt{N})$ steps to search a list with $N$ items; a classical
algorithm requires $\mathcal{O}\left(N\right)$ steps.
In our implementation of Grover's algorithm, the logical qubit basis states
$|00\rangle_L$, $|01\rangle_L$, $|10\rangle_L$ and $|11\rangle_L$ correspond to
the four items in our list.  We choose state $|11\rangle_L$ to correspond to
the item we wish to retrieve.  The algorithm's first step is to prepare the
register in an equal superposition of its basis states,
$|\psi\rangle_L = (1/2)\sum_{x_1=0}^1\sum_{x_2=0}^1|x_1x_2\rangle_L$.
The rest of the algorithm is an iterative process that increases the amplitude
of the sought-after state ($|11\rangle_L$) until it is the dominant part of the
superposition.  For our two-qubit register, only one iteration is required.
Finally, the logical qubits are measured; an outcome of $|11\rangle_L$
indicates that the algorithm has been successful.  (We have also used our
computer to implement the improved Deutsch-Jozsa algorithm described by Collins
\emph{et al.}~\cite{CKH98}, with similar results.)

\begin{figure}
\includegraphics{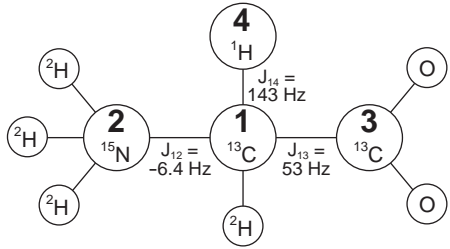}
\caption{\label{molfig}Isotope-substituted glycine molecule.  Spin-1/2
nuclei used as physical qubits are numbered 1--4.  The chemical shifts of
$\mbox{}^{13}\text{C}$ spins 1 and 3 differ by 16.5~kHz on a
500~MHz ($\mbox{}^1\text{H}$ resonance frequency) NMR spectrometer.}
\end{figure}

Our computer is based on the isotope-substituted glycine molecule
shown in Fig.~\ref{molfig}~\cite{comment3}.
Each of the molecule's four spin-1/2 nuclei
is used as a physical qubit.  The errors of Eq.~(\ref{errors})
are applied artificially according
to the following protocol.  At each of nine points in the experiment (indicated
in Fig.~\ref{grovfig}), error operator $X_1X_2I_3I_4$ is applied with
probability $e$, then
error operator $I_1I_2X_3X_4$ is applied with the same probability.  To
simulate the effect of a microscopic random process, the experiment is
performed 2048 times, each time with different randomly selected errors, and
the results are averaged,
giving the overall decoherence process a non-unitary, deterministic character.
The resulting decoherence increases with the error
probability $e$, becoming strongest at $e = 0.5$.  Formally, the operators
describing our engineered decoherence are $E_0 = \left(1-e\right)I_1I_2I_3I_4$,
$E_1 = \sqrt{e\left(1-e\right)}X_1X_2I_3I_4$,
$E_2 = \sqrt{e\left(1-e\right)}I_1I_2X_3X_4$, and $E_3 = eX_1X_2X_3X_4$.
(Note that for these error operators, it is clear that
$\sum_dE_d\rho_{|00\rangle_L}E_d^\dagger = \rho_{|00\rangle_L}$, where
$\rho_{|00\rangle_L}$ is defined by Eq.~(\ref{rhocomp}).)

\begin{figure*}
\includegraphics{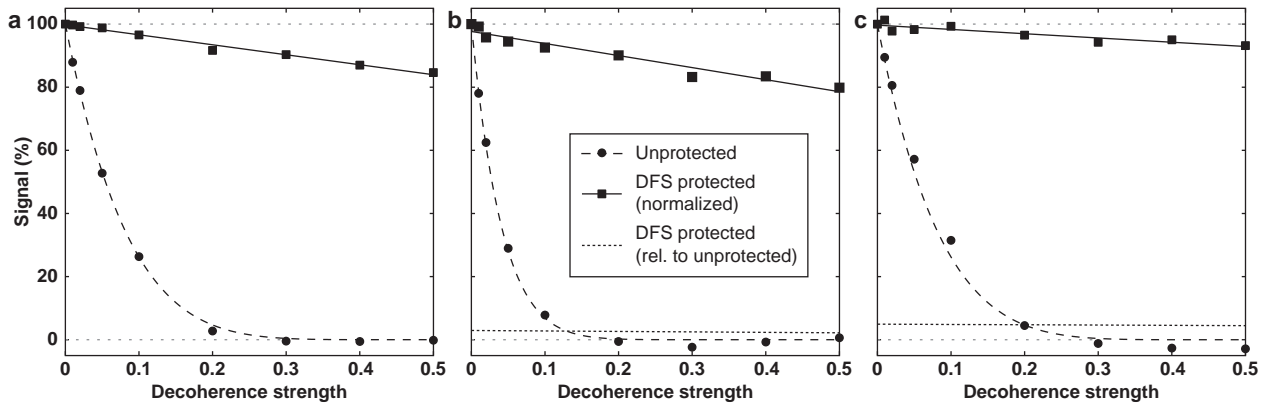}
\caption{\label{datafig}Experimental results in the presence of decoherence,
for both the error-avoiding and unprotected computers running Grover's
algorithm.
Results from the three temporal averaging steps appear separately in
(a,~$Z_1Z_2I_3I_4$), (b,~$Z_1Z_2Z_3Z_4$), and (c,~$I_1I_2Z_3Z_4$).
The integrated absolute intensity of the output signal measured at the end of
Grover's algorithm is charted as a function of decoherence strength $e$.
For some values of $e$, the signal from the unprotected computer is phase
inverted with respect to the correct output signal: In such cases we report a
negative signal intensity.
Square and round points are
experimental data from the error-avoiding and unprotected computers,
respectively, with each dataset normalized to the intensity of its first point.
Solid lines are linear regressions.
Dashed curves are a theoretically predicted intensity function,
$S = \left(1 - 2e\right)^n$ with $n = 6$ in (a, c) and $n = 12$ in (b),
obtained by predicting the unprotected computer's state at each of the
experiment's nine decoherence points and counting the number $n$ of error
operators which could change the system's state.
Dotted lines are signal
intensities from the error-avoiding computer normalized to the intensity of the
unprotected computer's $e = 0$ signal; the signals in (a) were recorded on
different nuclei and cannot be compared in this way.}
\end{figure*}

We have repeated the Grover algorithm experiment in the presence of nine
different levels of engineered decoherence ranging from $e = 0$ to $e = 0.5$.
At all values of $e$, the resulting NMR spectra contain little distortion and
clearly describe the final state of the qubit register as $|11\rangle_L$,
indicating the computer has successfully executed the algorithm.
To quantify the resistance of each computation to the
applied decoherence, we measured the integrated absolute intensity of the final
signal relative to the $e = 0$ experiment.  The dependence of signal intensity
on $e$ is different for each of the experiment's three temporal averaging steps
and we have chosen to analyze the results of the steps
separately (Fig.~\ref{datafig}).
We observe
only small losses of signal, and these cannot be attributed to any fault in the
DFS encoding.  The losses are predominantly due to imperfections in pulses used
to implement the engineered decoherence,
evidenced by a linear decrease in signal with increasing $e$.

As a control, we have repeated these experiments on a quantum computer that
does not use error avoidance.  The unprotected computer is similar to the
error-avoiding computer, with only the following changes: Spins 1 and 4
(Fig.~\ref{molfig}) serve directly as qubits (in place of the two logical
qubits), the
temporal averaging scheme prepares a different
(pseudo-pure)
initial state, and different
pulse sequences are used to implement quantum logic gates.  The same engineered
decoherence is applied and the computer executes the same algorithm as in the
error-avoiding computing experiments.  As expected, the unprotected computer's
signal intensity decreases rapidly with the error strength $e$.  We observe
effectively no signal when $e \geq 0.3$ and the result of the algorithm is
incorrect or unreadable for $e \geq 0.2$.

The DFS encoding our error-avoiding computer uses has an overhead cost, but the
results show it is small compared to the protection it affords.  The pulse
sequences that perform logical operations on the error-avoiding computer are
more complex than for the unprotected control computer, so the error-avoiding
computer is more vulnerable to signal loss due to pulse imperfections and
natural spin relaxation.  This is why, when signal intensities from the two
computers are compared on an absolute scale (dotted and dashed lines in
Fig.~\ref{datafig}),
the unprotected computer gives the stronger signal for low values of $e$.
However, the unprotected computer consistently fails when $e \geq 0.2$, while
the error-avoiding computer gives the correct result for all $e$.  The overall
fidelity of the Grover algorithm is governed by the temporal averaging step
that is least tolerant to decoherence (Fig.~\ref{datafig}b), and for this
step the
error-avoiding computer's signal is the more intense for
$e \geq$~$\sim\!\!0.15$.
Even at this low level of decoherence, the protection afforded by DFS encoding
outweighs the overhead involved.

In summary, we have provided the first experimental demonstration of quantum
computation in the presence of strong decoherence~\cite{comment1}, thus proving
that quantum error avoidance based on DFSs can
very effectively protect qubits from decoherence during the execution of a
quantum algorithm.  We have implemented Grover's search algorithm on two
two-qubit quantum computers, one error-avoiding and one unprotected.  While the
unprotected computer fails when exposed to even a moderate amount of
decoherence, the error-avoiding computer is successful in the presence of the
strongest possible decoherence.  This demonstration is a proof of the concept
of quantum error avoidance and suggests that DFS encoding will play an
important role in future experimental implementations of quantum algorithms in
the presence of decoherence.

\begin{acknowledgments}
We thank D.~R.~Muhandiram for assistance with NMR experiments and O.~Millet for
help in sample synthesis.  This work is supported by NSERC (all authors) and by
the DARPA-QuIST program, managed by AFOSR under agreement No. F49620-01-1-0468
(D.~A.~L.).  L.~E.~K. holds a Canada Research Chair in Biochemistry.
\end{acknowledgments}


\begin{thebibliography}{19}
\bibitem[{com({\natexlab{b}})}]{comment2}
\bibinfo{note}{We let $X$, $Y$, $Z$ represent the Pauli matrices $\sigma^x$,
  $\sigma^y$, $\sigma^z$ and drop the tensor product symbol $\otimes$.}
\bibitem[{com({\natexlab{c}})}]{comment3}
\bibinfo{note}{We prepared the isotope-substituted glycine by treating
  $[\mbox{}^{13}\text{C}$, $\mbox{}^{15}\text{N}$, $\mbox{}^1\text{H}]$-glycine
  with the enzyme glutamic-pyruvic transaminase in the presence of
  $\mbox{}^2\text{H}_2\text{O}$. (All reagents are commercially available from
  Sigma-Aldrich.) The sample we use in our NMR experiments contains
  $\sim\!\!50$ mg of this glycine in 500 $\mu$l $\mbox{}^2\text{H}_2\text{O}$.}
\end{thebibliography}
\end{document}